\documentclass[10pt]{iopart}
\usepackage{iopams,setstack,amsthm,mathrsfs,bm}
\usepackage[pdftex]{hyperref}

\begin{document}

\theoremstyle{definition}\newtheorem{lemma}{Lemma}
\theoremstyle{definition}\newtheorem{theorem}{Theorem}
\newtheorem{corollary}{Corollary}

\newcommand{\be}{\begin{equation}}
\newcommand{\ee}{\end{equation}}
\def\bgamma{\bar{\gamma}}
\title{Hilbert Spaces from Path Integrals}
\author{Fay Dowker$^1$, Steven Johnston$^1$, Rafael D. Sorkin$^{2,3}$}

\address{$^1$ Theoretical Physics, Blackett Laboratory, Imperial College London, London, SW7 2AZ, UK}
\address{$^2$ Perimeter Institute, 31 Caroline Street North, Waterloo ON, N2L 2Y5 Canada}
\address{$^3$ Department of Physics, Syracuse University, Syracuse, NY 13244-1130, USA}

\eads{\mailto{f.dowker@imperial.ac.uk}, \mailto{steven.johnston02@imperial.ac.uk} and \mailto{rsorkin@perimeterinstitute.ca}}

\font\german=eufm10 at 10pt \def\Buchstabe#1{{\hbox{\german #1}}}
\def\EA{\Buchstabe{A}} \def\IP{\langle \cdot,\cdot \rangle}
\def\L2{L^2(\mathbb{R}^d)}

\def\journaldata#1#2#3#4{{\it #1\/}\phantom{--}{\bf #2$\,$:} $\!$#3 (#4)}
\def\eprint#1{{\tt #1}}
\def\hpf#1{\webhome{\tt{some.papers/}}}
\def\webhome{{\tt http://www.perimeterinstitute.ca/personal/rsorkin/}}
\def\linebreak{\hfil\break}

\eqnobysec

\begin{abstract}
It is shown that a Hilbert space can be constructed for a quantum system
starting from a framework in which histories are fundamental.
The Decoherence Functional provides the inner product on this ``History
Hilbert space''.
It is also shown that the History Hilbert space {\emph{is}} the standard
Hilbert space in the case of non-relativistic quantum mechanics.
\end{abstract}


\section{Introduction}

It is not yet known how quantum theory and gravity will be reconciled.
However, the four-dimensional nature of reality revealed by our best
theory of gravity, General Relativity, suggests that unity in physics
will only be achieved if quantum theory can be founded on the concept of
{\emph{ history}} rather than that of {\emph{state}}.  The same
suggestion emerges even more emphatically from the causal set programme,
whose characteristic kind of spatio-temporal discreteness militates
strongly against any dynamics resting on the idea of Hamiltonian evolution.

A major step toward a histories-based formulation of quantum mechanics
was taken by Dirac and Feynman, showing that the quantum-mechanical
propagator can be expressed as a {\emph{sum over histories}}
\cite{Dirac:1933,Feynman:1948,FeynmanHibbs},
but it remains a challenge to make histories the foundational basis of
quantum mechanics.  One attempt to do this was made by J. Hartle who
set out new, histories-based axioms for
{\emph{Generalised Quantum Mechanics}} (GQM) which do not require the
existence of a Hilbert space of states \cite{Hartle1, Hartle2}.
Closely related in its technical
aspects --- whilst differing in interpretational aspiration --- is
{\emph{Quantum Measure Theory}} (QMT)
\cite{QuantumMeasure1, QuantumMeasure2, QuantMeasureInterp1, QuantMeasureInterp2}.
Thus far, both these approaches appear in the literature more as formal
axiomatic systems than as fully fledged mathematical physics,
although some concrete examples going beyond ordinary quantum mechanics
have been studied \cite{SorkinWalk}.

In this paper we take a step toward establishing QMT
and GQM more firmly on their foundations and
connecting them up with the more familiar formalism of state-vectors
and operators.
First we demonstrate in detail the Gel'fand-Naimark-Segal (GNS) type
construction given in \cite{SorkinWalk} of a
{\it History Hilbert space}
for any quantum measure system (to be defined).
It is technically helpful within quantum measure theory
that such a construction is available,
but the conceptual significance of this fact would be slight,
were it not that the constructed Hilbert space provably {\emph{is}} the
usual Hilbert space in the case of certain familiar quantum systems
(via an isomorphism that obtains formally in any unitary quantum theory
with pure initial state).
%
In this paper we exhibit non-relativistic particle quantum mechanics in
$d$ spatial dimensions as a quantum measure system, and we prove that
the Hilbert space constructed from the quantum measure is the usual
Hilbert space of (equivalence classes of) square integrable complex
functions on ${\mathbb{R}}^d$, given certain conditions on the
propagator.
The class of systems for which these conditions can be
established is large and includes the free particle and the simple
harmonic oscillator.
Thus, one of the main ingredients of text-book Copenhagen Quantum
Mechanics is derivable from the starting point of histories.

\section{Quantum Measure Theory: a histories-based framework}
 \label{HistoriesIntro}

We describe here the framework set out in \cite{QuantumMeasure1,
QuantumMeasure2, QuantMeasureInterp1, QuantMeasureInterp2}.  In
QMT, a physical, quantum system is associated with
a \emph{sample space} $\Omega$ of possible \emph{histories},
the space over which the integration of the path integral takes place.
Each history $\gamma$ in the sample space represents as complete a
description of physical reality as is
classically
conceivable in the theory.
The kind of elements in $\Omega$ varies from theory to theory.
In $n$-particle quantum mechanics, a history is a set of $n$
trajectories.  In a scalar field theory, a history is a real or complex
function on spacetime.
The business of discovering the appropriate sample space
for a particular theory is part of physics. Even in the seemingly
simple case of non-relativistic particle quantum mechanics, we do not
yet know what properties the trajectories in $\Omega$ should possess,
not to mention the knotty problems involved in defining $\Omega$ for
fermionic field theories for example. We will be able to sidestep
these issues in the current work.

\subsection{Event Algebra}

Once the sample space has been settled upon, any proposition about
physical reality is represented by a subset of $\Omega$. For example
in the case of the non-relativistic particle, if $R$ is a region of
space and $T$ a time, the proposition ``the particle is in $R$ at time
$T$'' corresponds to the set of all trajectories which pass through
$R$ at $T$.  We follow the standard terminology of stochastic
processes and refer to such subsets of $\Omega$ as \emph{events}.

An \emph{event algebra} on a sample space $\Omega$ is a non-empty
collection, $\EA$, of subsets of $\Omega$ such that
\begin{enumerate}
\item For any $\alpha \in \EA$, we have $\Omega\setminus \alpha \in
\EA$.
\item For any $\alpha, \beta \in \EA$, we have $\alpha \cup \beta \in
\EA$.
\end{enumerate}
An event algebra is then an algebra of sets
\cite{Halmos}. It
follows immediately that $\varnothing \in \EA$, $\Omega \in \EA$
($\varnothing$ is the empty set) and
$\EA$ is closed under finite unions and intersections.

An event algebra $\EA$ is a Boolean algebra under intersection
(logical ``and''), union (logical ``or'') and complement (logical
``not'') with unit element $\Omega$ and
zero element
$\varnothing$. It is also a (unital) ring with identity element
$\Omega$, multiplication as intersection and addition as symmetric
difference (logical ``xor''):
\begin{enumerate}
\item $\alpha \cdot \beta := \alpha \cap \beta$.
\item $\alpha + \beta := (\alpha \setminus \beta) \cup (\beta \setminus
\alpha).$
\end{enumerate}
This ring is Boolean since $\alpha \cdot \alpha =
\alpha$. It is also an algebra over $\mathbb{Z}_2$.
More discussion of the event algebra is given in
\cite{QuantMeasureInterp2}.

An example of an event algebra is the power set $2^\Omega:=\{ S : S 
\subseteq \Omega\}$ of all subsets of $\Omega$.
For physical systems with an infinite sample space,
however,
the event algebra will be strictly contained in the power set of $\Omega$,
something which is familiar from classical measure theory\footnote{%
To contrast with \emph{quantum} measure theory, the usual textbook
measure theory (see Halmos, \cite{Halmos}) will be called
``classical''.}
where the collection of ``measurable sets'' is not the whole power set.

If $\EA$ is also closed under countable unions and intersections then
$\EA$ is a $\sigma${\emph{-algebra}}.

\subsection{Decoherence Functional} \label{DecoherenceFunctional}

A \emph{decoherence functional} on an event algebra $\EA$ is a map $D
: \EA \times \EA \to \mathbb{C}$ such that
\begin{enumerate}
\item For all $\alpha, \beta \in \EA$, we have $D(\alpha,\beta) =
D(\beta,\alpha)^*$ (\emph{Hermiticity}).
\item For all $\alpha, \beta, \gamma \in \EA$ with $\beta \cap \gamma
= \varnothing$, we have $D(\alpha, \beta \cup \gamma) = D(\alpha,
\beta) + D(\alpha,\gamma)$ (\emph{Linearity}).
\item $D(\Omega, \Omega)=1$ (\emph{Normalisation}).
\item For any finite collection of events $\alpha_i \in \EA$
($i=1,\ldots,N$) the $N \times N$ matrix $D(\alpha_i,\alpha_j)$ is
positive semidefinite (\emph{Strong positivity}).
\end{enumerate}
A decoherence functional $D$ satisfying the weaker condition
$D(\alpha,\alpha) \geq 0$ for all $\alpha \in \EA$ is called
\emph{positive}.  Note that in Generalised Quantum Mechanics,
a decoherence functional is defined to be positive
rather than strongly positive \cite{Hartle1, Hartle2}.

A \emph{quantal measure} on an event algebra $\EA$
is a map $\mu: \EA \to \mathbb{R}$ such that
\begin{enumerate}
\item For all $\alpha \in \EA$, we have $\mu(\alpha) \geq 0$
(\emph{Positivity}).
\item For all mutually disjoint $\alpha, \beta, \gamma \in \EA$, we
have
\begin{equation*}
\fl \mu(\alpha \cup \beta \cup \gamma) - \mu(\alpha \cup \beta) -
\mu(\beta \cup \gamma) - \mu(\alpha \cup \gamma) +\mu(\alpha) +
\mu(\beta) + \mu(\gamma) = 0\,.
\end{equation*}
({\emph{Quantal Sum Rule}})
\item $\mu(\Omega)=1$ (\emph{Normalisation}).
\end{enumerate}

If $D : \EA \times \EA \to \mathbb{C}$ is a decoherence functional
then the map $\mu : \EA \to \mathbb{R}$ defined by
$\mu(\alpha):=D(\alpha,\alpha)$ is a quantal measure.

A triple, $(\Omega, \EA, D)$, of sample space, event algebra and
decoherence functional will be called a {\emph{quantum measure system}}.

%

\subsection{A Hilbert Space Construction}

Given a quantum measure system, $(\Omega, \EA, D)$, we can construct a
Hilbert space: a complex vector space with (non-degenerate) Hermitian
inner product which is complete with respect to the induced norm.
This construction is given in \cite{SorkinWalk} and is
essentially that given by V.P. Belavkin in
\cite[Theorem 3, Part 1]{Belavkin} where the decoherence functional is
called a ``correlation kernel''.
The construction is akin to the GNS construction
of a Hilbert space from a $C^*$-algebra and is the same as the
construction appearing in
Kolmogorov's Dilation Theorem \cite[Theorem 2.2]{Dutkay}, \cite{EvansLewis}.

To start, we first construct the free vector space on $\EA$ and use the
decoherence functional to define a degenerate inner product on it.

\subsubsection{Inner product space: $H_1$} \label{sec:h1}

To define the \emph{free vector space} on an event algebra $\EA$ we
start with the set of all complex-valued functions on $\EA$
which are non-zero only on a finite number of events. This set becomes
a vector space, $H_1$, if addition and scalar multiplication are
defined by:
\begin{enumerate}
\item For all $u, v \in H_1$ and $\alpha \in \EA$, we have
$(u+v)(\alpha) := u(\alpha) + v(\alpha)$.
\item For all $u \in H_1$, $\lambda \in \mathbb{C}$ and $\alpha \in
\EA$, we have $(\lambda u)(\alpha):=\lambda u(\alpha)$.
\end{enumerate}

We now define an inner product space $(H_1,\IP_1)$ by defining a
degenerate inner product on $H_1$ using the decoherence functional
$D$. For $u,v \in H_1$ define: \be \langle u, v \rangle_1 :=
\sum_{\alpha \in \EA}\sum_{\beta \in \EA} u(\alpha)^* D(\alpha,\beta)
v(\beta).\ee This sum is well-defined because $u$ and $v$ are non-zero
for only a finite number of events. This satisfies the conditions
for an inner product. Note that the strong positivity of the
decoherence functional is essential for $\langle u, u \rangle_1 \geq 0$.

To see that the inner product is degenerate consider, for example, the
non-zero vector $u \in H_1$ defined by: \be u(x) :=
\left\{\begin{array}{rl} 1 & \textrm{if } x = \alpha, \\ 1 &
\textrm{if } x = \beta, \\ -1 & \textrm{if } x = \alpha \cup \beta, \\
0 & \textrm{otherwise}\end{array} \right.\ee  for
two nonempty, disjoint events $\alpha, \beta \in \EA$. By applying the
properties of the decoherence functional we see that $||u||_1 = 0$.

\subsubsection{Hilbert space: $H_2$}\label{sec:h2}

We now quotient and complete the inner product space $(H_1,\IP_1)$ to form a
Hilbert space $(H_2,\IP_2)$.

For two Cauchy sequences $\{u_n\},\{v_n\}$ in $H_1$ we define an
equivalence relation \be \label{eq:EquivRelation1} \{u_n\} \sim_1
\{v_n\} \iff \lim_{n\rightarrow \infty} ||u_n - v_n||_1 = 0.\ee We
denote the $\sim_1$ equivalence class of a Cauchy sequence $\{u_n\}$
by $[u_n]_1$. The set of these equivalence classes form a Hilbert
space, $(H_2,\IP_2)$, if addition, scalar multiplication and the inner
product are defined by:
\begin{enumerate}
\item For all $[u_n]_1, [v_n]_1 \in H_1$, we have $[u_n]_1 +
[v_n]_1:=[u_n + v_n]_1.$
\item For all $[u_n]_1 \in H_1$ and $\lambda \in \mathbb{C}$, we have
$\lambda[u_n]_1 :=[ \lambda u_n]_1.$
\item For all $[u_n]_1, [v_n]_1 \in H_1$, we have \be
\label{eq:HilbInnerProd2} \langle [u_n]_1, [v_n]_1 \rangle_2 :=
\lim_{n \rightarrow \infty} \langle u_n, v_n \rangle_1\ee
\end{enumerate}
These are all well-defined, independent of which representative is
chosen from the equivalence classes.

The construction of a Hilbert space (here $(H_2,\IP_2)$) from an inner
product space (here $(H_1,\IP_1)$)
is a standard
operation described in many textbooks
(for example, \cite[Section 7]{Geroch}, \cite[p198]{Binmore}).

Whether or not $H_2$ is separable depends on the particular event
algebra and decoherence functional that are used in its
construction\footnote{The dimension of $H_1$ is equal to the
cardinality of $\EA$ but the dimension of $H_2$, which is less than
that of $H_1$, depends on the $\sim_1$ equivalence relation (which in
turn depends on $D$).}.  In Sections \ref{FiniteSampleSpace},
\ref{Particle} and \ref{ParticlePotential} we shall present systems
for which the constructed Hilbert space is isomorphic to a separable
Hilbert space (the standard Hilbert space for the system). In these
examples the constructed Hilbert space is therefore separable.

Note that we did not use the full structure of the quantum measure
system: only the event algebra, $\EA$ and the decoherence functional
$D$ were used and nowhere did the underlying sample space enter into
the game.  This will be important in our discussion of particle
quantum mechanics where there is an event algebra $\EA$ but we have no
precise definition, as yet, of the sample space.

We will refer to the Hilbert space, $H_2$, constructed from a quantum
measure system as the {\emph{History Hilbert space}}.
For quantum systems which have a standard, Copenhagen formulation in
terms of unitary evolution on a Hilbert space of states
%
and which can {\emph {also}} be cast
into the form of a quantum measure system, the question arises
as to the
relationship between the standard Hilbert space and the History
Hilbert space.
This is the question under study in this paper and
it will be shown that in general the answer depends on the
initial state and the Schr\"odinger dynamics for the system
since these are what define the decoherence functional.
However, we conjecture that {\emph{generically}}
where both Hilbert
spaces exist and the decoherence
functional encodes a pure initial state,
they are isomorphic. Moreover the isomorphism is
physically meaningful, so that one can conclude that the History
Hilbert space {\emph{is}} the standard Hilbert space of the system.

We will prove this conjecture for a variety of non-relativistic
particle systems and exhibit the isomorphism explicitly.
The systems considered include a particle with a
finite configuration space, a free non-relativistic particle in $d$ spatial
dimensions, and a non-relativistic particle in various backgrounds,
including a quadratic potential and an infinite potential barrier.

Before turning to these specific cases, we recall the following simple
lemma.
\begin{lemma} \label{lem:OneToOne} A linear map $f : H_A \to H_B$
 from a Hilbert space $(H_A,\IP_A)$ to a Hilbert space $(H_B, \IP_B)$
that preserves the inner product, {\emph{i.e.}}
\be \langle f(u), f(v) \rangle_B = \langle u, v \rangle_A \ee for all
$u,v \in H_A$, is one-to-one.
\begin{proof}
For all $u,v \in H_A$ we have 
\be \fl f(v) = f(u) \iff 0 = ||f(u) -
f(v)||_B = ||f(u-v)||_B = ||u - v||_A \iff u = v \ee
\end{proof}
\end{lemma}

\section{Finite Configuration Space} \label{FiniteSampleSpace}

We analyse the case of a unitary quantum system with finite
configuration space as a warm up for the system of main interest,
particle quantum mechanics.
Consider a system which has a finite configuration space of $n$
possible configurations at any time. We shall only consider the system's
configuration
at a finite number $N$ of fixed times
$t_1 = 0 < t_2 < \ldots < t_N = T$.
An example of such a system is a particle with $n$
possible positions at each time which evolves in $N-1$ discrete
time-steps from time $t=0$ to time $t=T$.

\subsection{Standard Hilbert space approach}

The Hilbert space for the system is $(\mathbb{C}^n,\IP)$
and states of the system
at a particular time are represented by vectors in $\mathbb{C}^n$. For
a state $\psi \in \mathbb{C}^n$ the $i^{\textrm{th}}$ component,
$\psi_i$, is the amplitude that the system is in configuration $i$.
 For all $\psi, \phi \in \mathbb{C}^n$
the non-degenerate inner product is given by \be \langle \psi,
\phi\rangle:= \sum_{i=1}^n \psi^*_i \phi_i. \ee

There exists a time evolution operator, $U(t',t)$,
a unitary transformation
which evolves states at time $t$ to states at time $t'$
and which satisfies the folding property
\be
U(t'',t')U(t',t) = U(t'', t)\,.
\ee

\subsection{A Quantum Measure System}

Each history, $\gamma$, of the system is represented by an $N$-tuple of integers
$\gamma = (\gamma_1,\gamma_2,\ldots,\gamma_N)$
(with $1 \leq \gamma_a \leq n$ for all
$a=1,\ldots,N$) where each integer $\gamma_a$ denotes the configuration of
the system at time $t=t_a$. The system's sample space, $\Omega$, is
the (finite) collection of these $n^N$ possible histories. The
event algebra, $\EA$, is the power set of $\Omega$: $\EA := 2^\Omega =
\{ S : S \subseteq \Omega\} $.

To define the decoherence functional we assume there is an initial
state $\psi \in \mathbb{C}^n$ of unit norm. This can be thought of
as a vector in $\mathbb{C}^n$ or simply as an $n$-tuple of amplitudes
weighting each initial configuration at time $t=0$.
The decoherence functional for singleton events is,
\begin{eqnarray}
 \nonumber D(\{\gamma\}, \{\bgamma\}) &:= \psi(\gamma_1)^* U_{\gamma_2 \gamma_1}^* U_{\gamma_3\gamma_2}^*
 \dots U_{\gamma_N\gamma_{N-1}}^*{~} \\ & \delta_{\gamma_N \bgamma_N}
 U_{\bgamma_N\bgamma_{N-1}} \dots U_{\bgamma_2 \bgamma_1} \psi(\bgamma_1)
 \label{dcfsingle.eq}
\end{eqnarray}
where $\gamma,\bgamma\in \Omega$, $\psi(\gamma_1)$ is the $\gamma_1$-th
component of $\psi$
and $U_{\gamma_2 \gamma_1}$ is short hand for
$U(t_2, t_1)_{\gamma_2 \gamma_1}$, the amplitude to go from
$\gamma_1$ at $t_1$ to $\gamma_2$ at $t_2$.
$D$ has ``Schwinger-Kel'dysh'' form, equalling the complex conjugated
amplitude of $\gamma$ times the amplitude of ${\bgamma}$ when
the two histories end at the same final position, and zero otherwise.
The decoherence functional of events $\alpha, \beta \in \EA$
is then fixed by the bi-additivity property:
\be
  D(\alpha,\beta): = \sum_{\gamma\in \alpha} \sum_{\bgamma\in\beta}
  D(\{\gamma\}, \{\bgamma\})\,.
\ee

We define the {\emph{restricted evolution}}
of the initial state $\psi \in \mathbb{C}^n$
with respect to a history $\gamma$
to be the state $\psi_\gamma \in \mathbb{C}^n$
given by:
\be
  \psi_\gamma := P^{\gamma_N} U(t_N,t_{N-1}) P^{\gamma_{N-1}}\cdots
  P^{\gamma_3}U(t_3,t_2) P^{\gamma_2} U(t_2,t_1) P^{\gamma_1}\psi
\ee
where $P^i$ is the projection operator in $\mathbb{C}^n$ that projects
onto the state which is non-zero only on the $i$th configuration.
[Thus $\psi_\gamma$ is just the configuration $\gamma_N$ weighted by the
 amplitude $U_{\gamma_N\gamma_{N-1}}\dots U_{\gamma_2 \gamma_1}\psi(\gamma_1)$.]
Restricted evolution of the initial state with respect to an event
$\alpha$ is then defined to be the state $\psi_\alpha \in \mathbb{C}^n$
\be
\psi_\alpha := \sum_{\gamma \in \alpha} \psi_\gamma. \ee
Note that $\psi_\gamma = \psi_{\{\gamma\}}$, so we can use
either notation when an event is a singleton.
It is easy to see that
the decoherence functional for two events $\alpha, \beta \in \EA$ is
equal to the inner product between the two restricted
evolution states, $\psi_\alpha$ and $\psi_\beta$:
\be\label{dcf} D(\alpha,\beta):= \langle \psi_\alpha, \psi_\beta \rangle. \ee

\subsection{Isomorphism}

We now look at conditions on the initial state and evolution of the
system that ensure the History Hilbert space $(H_2,\IP_2)$ is
isomorphic to $(\mathbb{C}^n, \IP)$.

For this system both the sample space and event algebra are
finite
so the inner product space $(H_1,\IP_1)$ is finite dimensional and
therefore complete but with a degenerate inner product. In this
case there is no need to consider Cauchy sequences of
elements of $H_1$. Instead, we define the equivalence relation
directly on $H_1$: $u\sim_1 v$ if $||u - v||_1 = 0$. And
$H_2$ is defined as $H_2 := H_1/\sim_1$ the
space of equivalence classes, $[u]_1$ under $\sim_1$.
For all $u, v \in H_1$, we have by (\ref{dcf})
\be
\label{eq:FiniteInnerProdContinuity} \langle [u]_1,[v]_1
\rangle_2 := \langle u, v
\rangle_1.
\ee

It will prove useful to define a map $f_0 : H_1 \to \mathbb{C}^n$
given by \be f_0(u):= \sum_{\alpha \in \EA} u(\alpha) \psi_{\alpha},
\ee for all $u \in H_1$. This sum is well-defined since $u(\alpha)$ is
non-zero for only a finite number of $\alpha \in \EA$. This $f_0$ is linear and,
for all $u, v \in H_1$, we have \be
\label{eq:FiniteInnerprod} \langle f_0(u), f_0(v) \rangle = \langle u,
v \rangle_1, \ee which ensures
\be \label{eq:FiniteEquiv} [u ]_1 = [v]_1
\Rightarrow f_0(u) = f_0(v). \ee


Using the map $f_0$ we define the candidate isomorphism $f : H_2 \to
\mathbb{C}^n$ by \be \label{eq:FiniteFMap} f([u]_1) := f_0(u),\ee for
all $[u]_1 \in H_2$. By \eref{eq:FiniteEquiv}, $f$ is well-defined,
independent of the equivalence class representative chosen.
The map $f$ is linear and \eref{eq:FiniteInnerProdContinuity} and
\eref{eq:FiniteInnerprod} ensure that for all $[u]_1, [v]_1 \in H_2$,
we have: \be \label{eq:FiniteInnerProductPreserving}\langle f([u]_1),
f([v]_1) \rangle = \langle [u]_1, [v]_1 \rangle_2.\ee


By Lemma \ref{lem:OneToOne}, since $f$ is linear and satisfies
\eref{eq:FiniteInnerProductPreserving}, it is one-to-one. If we can
find a condition on the initial state and dynamics that ensures the
map $f$ is onto then it is the isomorphism we seek.

\begin{theorem}[Onto] \label{the:FiniteOnto}
Let the evolution operators $U(t',t)$ and initial state $\psi \in
\mathbb{C}^n$ be such that, for each configuration $j=1,\ldots,n$ at
the final time, there exists a history ending at $j$,
$\gamma^j = (\gamma^j_1,\gamma^j_2,\ldots,\gamma^j_{N-1},j) \in \Omega$,
with non-zero amplitude. In other words, the $j$-th component of
the restricted evolution of the initial state with respect to
history $\gamma^j$ is non-zero: $(\psi_{\gamma^j})_j \neq 0$. Then the map $f$ is onto.
\begin{proof}
For each $j$ choose a history $\gamma^j \in \EA$ such that $(\psi_{\gamma^j})_j \neq
0$ (note that $\psi_{\gamma^j}$ is only non-zero in the $j$-th component). Let $\phi \in \mathbb{C}^n$ be a vector we wish to map to.

Define $u \in H_1$ by \be u(x) := \left\{\begin{array}{rl}
\phi_j/(\psi_{\gamma^j})_j & \textrm{if } x = \{\gamma^j\} \textrm{ for }
j=1,\ldots,n, \\ 0 & \textrm{otherwise.}\end{array} \right.\ee This is
a well-defined vector in $H_1$ and satisfies $f([u]_1) = \phi$.  Hence
$f$ is onto.
\end{proof}
\end{theorem}

An example of a case in which $H_2$ is not isomorphic to $\mathbb{C}^n$
is if the initial state has support only on a single
configuration, $k$,  and the evolution is trivial, $U(t,t') = 1$.
Then the only configuration at the final time with nonzero
amplitude is $k$ and the History Hilbert space is one dimensional,
not  $\mathbb{C}^n$. Another example is if the evolution is ``local'' on the
lattice, so that after the first time step, only $k$ and $k\pm 1$ say
have nonzero amplitude. Then the dimension of the History Hilbert space
will depend on the number of time steps and will grow with $N$ until it
reaches $n$ after which it will be constant.

\section{Particle in $d$ dimensions} \label{Particle}

We turn now to a less trivial system, that of
a non-relativistic particle moving in $d$ dimensions.

\subsection{Hilbert space approach}

We recall some basic technology in order to
fix our notation. The Hilbert space for the system
is $(\L2,\IP)$. In order to define this,
we first define the inner product space
$(\mathcal{L}^2(\mathbb{R}^d),\IP_0)$, the space of square
integrable
functions $\psi : \mathbb{R}^d \to
\mathbb{C}$.
For all $\psi,\phi \in
\mathcal{L}^2(\mathbb{R}^d)$ a degenerate inner product is given by
\be \langle \psi,\phi\rangle_0 := \int_{\mathbb{R}^d}
\psi^*(\mathbf{x}) \phi(\mathbf{x}) d\mathbf{x}. \ee To see that the
inner product is degenerate consider any vector $\psi \in
\mathcal{L}^2(\mathbb{R}^d)$ which is non-zero only on a set of
measure zero. Although $\psi \neq 0$, we have $||\psi||_0
= 0$.

For two vectors $\psi, \phi \in \mathcal{L}^2(\mathbb{R}^d)$
define the equivalence relation $\sim$ by
\be \label{eq:ParticleEquivDef}
\psi \sim \phi \iff ||\psi - \phi||_0 = 0.\ee
The $\sim$ equivalence
class of $\psi \in \mathcal{L}^2(\mathbb{R}^d)$ will be denoted by
$[\psi]$. The set of all equivalence classes forms the Hilbert space
$(\L2,\IP)$
where, for all $[\psi], [\phi] \in \L2$, $\langle
[\psi],[\phi]\rangle := \langle \psi,\phi\rangle_0$.
State vectors for the particle at a fixed time are vectors
in $\L2$.

\subsection{Quantum Measure System}

The sample space of the system, $\Omega$, is the set of all
continuous\footnote{We choose continuous maps for definiteness but
recognise that the correct sample space may
have more refined continuity conditions or even be something more
general. The results of our work will
remain applicable so long as the actual event algebra contains
a subalgebra isomorphic to the $\EA$ we define here and
on which the measure is defined by the propagator
in the same --- standard --- way.}
maps $\gamma : [0,T] \to \mathbb{R}^d$. These maps represent the
trajectory of the particle from an initial time $t=0$ to a final
``truncation time'' $t=T$.

Introducing a truncation time $T$ seems necessary for the construction
undertaken below, which produces the quantal measure for the
corresponding subalgebra $\EA_T\subseteq\EA$.  This limitation to a
subalgebra of the full event algebra is only apparent, however, because
$\EA$ is the union of the $\EA_T$, and the measure of an event $A\in\EA$
does not depend on which subalgebra we refer it to.  In section
\ref{InfTime} we explain this in detail for the case of unitary theories
such as we are concerned with in the present paper.


\subsubsection{Event algebra}

The event algebra $\EA$ we now define is strictly contained in the power
set $2^\Omega$.
Let $N$ be any positive integer, $N \ge 2$.
Let $\mathbf{t} = (t_1, t_2, \ldots, t_N)$ be
any $N$-tuple of real numbers with $0 = t_1 < t_2 < \ldots < t_N = T$
and $\bm{\alpha} = (\alpha_1,\alpha_2,\ldots,\alpha_N)$
any $N$-tuple of subsets of $\mathbb{R}^d$ such that,
for each $k=1,\ldots,N$,
either $\alpha_k$ or its complement $\alpha_k^c$ is a bounded Lebesgue measurable set.
A subset $\alpha \subseteq \Omega$ is called a
\emph{homogeneous event}\footnote%
{Alternative names include elementary event, regular event or cylinder set.}
\cite{Isham}
if there exists an integer $N$ and a pair
$(\textbf{t}, \bm{\alpha})$ such that
\be \label{eq:HomoDefn}
 \alpha = \{ \gamma \in \Omega : \gamma(t_k) \in \alpha_k, k=1,\ldots,N \}.
\ee
Each $\alpha_k$
can be thought of as a condition on the system, a restriction on
the position of the particle, at time $t_k$.
We represent a homogeneous event
by the pair $(\textbf{t},\bm{\alpha})$. This representation is
non-unique because, for example, the same
homogeneous event $\alpha$ is represented by the pairs
\be \textbf{t} := (t_1,t_2,t_3), \quad \bm{\alpha}:=
(\alpha_1,\alpha_2,\alpha_3),\ee and \be \textbf{t}' :=
(t_1,t_2,t',t_3), \quad \bm{\alpha}' :=
(\alpha_1,\alpha_2,\mathbb{R}^d, \alpha_3).\ee

The event algebra $\EA$ is defined to be
the collection of all finite unions of
homogeneous events. Any event $\alpha \in \EA$
which is not a homogeneous event will be called \emph{inhomogeneous}.

We can better understand the structure of the event algebra if we
consider a few set operations in it.
Abusing notation slightly we'll represent a homogeneous event $\alpha$
(with representation $(\bm{t},\bm{\alpha})$) by its ordered collection
of sets:
$\alpha = (\alpha_1,\alpha_2,\ldots,\alpha_N)$.
The complement of $\alpha$ is then a finite union of $2^N-1$ disjoint
homogeneous events.
For example for $\alpha = (\alpha_1,\alpha_2,\alpha_3)$ we have
 \begin{eqnarray} \alpha^c &= (\alpha^c_1,\alpha_2,\alpha_3) \cup (\alpha_1,\alpha^c_2,\alpha_3) \cup (\alpha_1,\alpha_2,\alpha^c_3)\\&
  \nonumber \cup (\alpha^c_1,\alpha^c_2,\alpha_3) \cup (\alpha^c_1,\alpha_2,\alpha^c_3) \cup (\alpha_1,\alpha^c_2,\alpha^c_3) \cup (\alpha^c_1,\alpha^c_2,\alpha^c_3)\end{eqnarray}
where $^c$ denotes set-complement.

The intersection of two homogeneous events $\alpha = (\alpha_1,\alpha_2,\ldots,\alpha_N), \beta = (\beta_1,\beta_2,\ldots,\beta_N)$
(which, by adding extra copies of $\mathbb{R}^d$ as needed, can be assumed to have the same time-sequence $\bm{t}$)
is the homogeneous event $\alpha \cap \beta = (\alpha_1 \cap \beta_1,\alpha_2\cap \beta_2,\ldots,\alpha_N\cap \beta_N)$.

These two properties say that the homogeneous events form a
\emph{semiring} and ensure that for two homogeneous events $\alpha$
and $\beta$ the event $\alpha \setminus \beta = \alpha \cap \beta^c$
is a finite union of disjoint homogeneous events.
This means that a finite union of homogeneous events can be re-expressed
as a finite union of \emph{disjoint} homogeneous events.
As an example consider the event
$\alpha = \alpha_H^1 \cup \alpha_H^2 \cup \alpha_H^3$
for three homogeneous events
$\alpha_H^A$ ($A=1,2,3$). We can define three disjoint events
$\bar{\alpha}^A$ by
\be
\label{eq:HomogeneousDisjoint1}
\bar{\alpha}^1 =
\alpha_H^1,\quad \bar{\alpha}^2:=\alpha_H^2 \setminus
\alpha_H^1,\quad \bar{\alpha}^3:=(\alpha_H^3 \setminus
\alpha_H^1) \cap (\alpha_H^3 \setminus \alpha_H^2).\ee
Now, from the remarks above,
\be
 \alpha_H^2 \setminus \alpha_H^1 =
\bigcup_{i=1}^{N_1} \beta_1^i, \quad \alpha_H^3 \setminus \alpha_H^1
= \bigcup_{j=1}^{N_2} \beta_2^j, \quad
\alpha_H^3 \setminus \alpha_H^2 = \bigcup_{k=1}^{N_3} \beta_3^k,
\ee
where $\beta^i_1, \beta^j_2, \beta^k_3$ are homogeneous events
such that $\beta_A^i \cap \beta_A^{i'} = \varnothing$ if $i \neq i'$
(for $A=1,2,3$ and $i,i'=1,\ldots,N_A$).

We therefore have
\begin{eqnarray}
\label{eq:HomogeneousDisjoint2}
\alpha &= \bar{\alpha}^1 \cup
 \bar{\alpha}^2 \cup \bar{\alpha}^3 = \alpha_H^1
 \cup \left(\bigcup_{i=1}^{N_1} \beta_1^i\right) \cup
 \left(\left(\bigcup_{j=1}^{N_2} \beta_2^j \right) \cap
 \left( \bigcup_{k=1}^{N_3} \beta_3^k \right)\right)\\ &\nonumber =
 \alpha_H^1 \cup \left(\bigcup_{i=1}^{N_1}
 \beta_1^i\right) \cup \left(\bigcup_{j=1}^{N_2} \bigcup_{k=1}^{N_3}
 \beta_2^j \cap \beta_3^k \right)\end{eqnarray}
which expresses $\alpha$ as a finite union of \emph{mutually
disjoint} homogeneous events---namely $\alpha_H^1,
\beta_1^i$ ($i=1,\ldots,N_1$) and $\beta_2^j \cap \beta_3^k$
($j=1,\ldots,N_2, k=1,\ldots,N_3$). The procedure
followed in this example extends without difficulty to $M>3$
homogeneous events but with an associated
proliferation of notation.

(For representing such relationships, the Boolean-algebraic notation can
be quite expressive.  For example, the essence of
\eref{eq:HomogeneousDisjoint1}-\eref{eq:HomogeneousDisjoint2}
is the disjoint decomposition, for any three events,
$\alpha\cup\beta\cup\gamma=\alpha+(1+\alpha)\beta+(1+\alpha)(1+\beta)\gamma$.
Notice here that $1+\alpha$ is the complement of $\alpha$, as is clearly
visible in the calculation,
$\alpha\cap(1+\alpha)\equiv\alpha(1+\alpha)=\alpha+\alpha^2=\alpha+\alpha=0$.)

The event algebra $\EA$ defined here is an algebra but not a
$\sigma$-algebra. We allow only a finite number of times when defining
a homogeneous event which means $\EA$ is closed under finite
unions but not under countable unions. In
Section \ref{ParticleDecoherence}, a decoherence functional will be
defined on $\EA$. It is not clear whether this definition can be
\emph{extended} to define a decoherence functional on the
full $\sigma$-algebra (of subsets of $\Omega$) generated by $\EA$.
For this
to be done it would require a ``fundamental theorem of quantum
measure theory'' analogous to the Carath\'{e}odory-Kolmogorov Extension Theorem
for classical measures (Theorem A, p. 54 of \cite{Halmos})

\subsubsection{Decoherence functional} \label{ParticleDecoherence}

Let $\psi \in \mathcal{L}^2(\mathbb{R}^d)$ be the normalised initial
state, then the decoherence functional for singleton events is given
formally by
\be
  D(\{\gamma\}, \{\bgamma\}) := \,\, \psi(\gamma(0))^* e^{-i S[\gamma]}
  \delta(\gamma(T)-\bgamma(T)) e^{iS[\bgamma]} \psi(\bgamma(0)) \,.
\ee
By bi-additivity, the decoherence functional
for events $\alpha, \beta \in \EA$
is then given by the double path integral:
\be \label{dcf4qm}
  D(\alpha, \beta) := \,\,
  \int_{\gamma\in \alpha} [d\gamma]\int_{\bgamma \in \beta}[d\bgamma]
  D(\{\gamma\}, \{\bgamma\}) \,.
\ee
All these formulae are, as yet, only formal.
We do not
know rigorously
what $\Omega$ is,
whether the singleton subsets of $\Omega$ are measureable,
or how to define the integration-measure $[d\gamma]$.
Indeed, one might anticipate that, as with Wiener measure, neither
$e^{iS[\gamma]}$ nor $[d\gamma]$ can be defined separately, and only
their combination in  (\ref{dcf4qm}) will exist mathematically.

Nonetheless, we can make sense of the decoherence functional
(\ref{dcf4qm}) on $\EA$ because the form of the events --- unions of
homogeneous events --- allows us to equate the path integrals in
(\ref{dcf4qm}) to well-defined expressions involving the
\emph{propagator}. 
The propagator is a function\footnote{The propagator may in general be
a distribution, as in the case of a simple harmonic oscillator example
in Section \ref{sec:ParticleExamples}.}
$K(\mathbf{x}',t'|\mathbf{x},t)$ that encodes the dynamics of the
particle. We assume that the dynamics of the system is unitary.
We define the restricted evolution of $\psi \in
\mathcal{L}^2(\mathbb{R}^d)$ according to a homogeneous event $\alpha
\in \EA$ (with representation $\mathbf{t} = (t_1, t_2, \ldots, t_N)$
and $\bm{\alpha} = (\alpha_1,\alpha_2,\ldots,\alpha_N)$) to be
$\psi_\alpha$ given by

\begin{eqnarray} \psi_\alpha(\mathbf{x}_T,T) &:=
\chi_{\alpha_N}(\mathbf{x}_T) \int_{\alpha_{N-1}} \!\!\!\!\!\!\!
d\mathbf{x}_{N-1} \int_{\alpha_{N-2}} \!\!\!\!\!\!\! d\mathbf{x}_{N-2}
\cdots \int_{\alpha_2}\!\!\!\! d\mathbf{x}_2 \int_{\alpha_1} \!\!\!\!
d\mathbf{x}_1 \nonumber\\&
\quad K(\mathbf{x}_T,T|\mathbf{x}_{N-1},t_{N-1}) \cdots
K(\mathbf{x}_2,t_2|\mathbf{x}_1,0) \psi(\mathbf{x}_1)
\label{eq:RestrictedEvolHomo}, \end{eqnarray}
where
\be \chi_{A} (x) := \left\{\begin{array}{l} 1 \textrm{ if } x
\in A, \\ \\ 0 \textrm{ if } x \notin A.\end{array} \right.\ee is the
characteristic function of $A \subset \mathbb{R}^d$.

The convergence of the integrals (and therefore the existence of
$\psi_\alpha$) in \eref{eq:RestrictedEvolHomo} depends on the
propagator for the system
and the type of $\alpha_k$ subsets allowed.
For the examples we shall consider\footnote{These examples include the free particle
and the simple harmonic oscillator.} in Section \ref{sec:ParticleExamples}
the integrals converge if all the
$\alpha_k$ subsets are bounded and, it turns out,
in the isomorphism proof in Section \ref{ParticleIsomorphism} we will only require such
events. In fact we will deal only with two-time homogeneous events
with bounded measurable sets at the initial and final times.

Nevertheless we must still define the decoherence functional on the
entire event algebra $\EA$ and to do this we must define restricted
evolution according to a homogeneous event $\alpha$ when some of the
$\alpha_k$ subsets are
unbounded (which, for the event algebra we are considering, only happens
if the $\alpha_k$ are \emph{complements} of bounded measurable sets).
In general (and certainly for the examples we shall look at) the propagator is oscillatory in position
and if $\alpha_k$, say,
is unbounded
the $d\mathbf{x}_k$
integral in \eref{eq:RestrictedEvolHomo} does not converge absolutely.

We deal with this non-convergence in the standard way (see
{\emph{e.g.}}
\cite[footnote 13]{Feynman:1948})
by introducing a convergence factor.
For each
unbounded
$\alpha_k$ we
replace the
non-convergent
$d\mathbf{x}_k$ integral \be
\label{eq:IntNoConvergenceFactor} \int_{\alpha_k}
K(\mathbf{x}_{k+1},t_{k+1}|\mathbf{x}_k,t_k)
K(\mathbf{x}_k,t_k|\mathbf{x}_{k-1},t_{k-1})d \mathbf{x}_k, \ee in
\eref{eq:RestrictedEvolHomo} by \be \label{eq:IntConvergenceFactor}
\lim_{\epsilon \to 0^+} \int_{\alpha_k}
K(\mathbf{x}_{k+1},t_{k+1}|\mathbf{x}_k,t_k)
K(\mathbf{x}_k,t_k|\mathbf{x}_{k-1},t_{k-1}) \exp\left( - \epsilon
\mathbf{x}_k^2 \right)d \mathbf{x}_k.\ee For the propagators we
consider this integral converges and the $\epsilon \to 0^+$ limit
exists.
By using these convergence factors
we can define $\psi_\alpha$ for all homogeneous events $\alpha
\in \EA$.

For the propagators we will consider, the following composition
property holds: \be \label{eq:ESCK} \lim_{\epsilon \to 0^+}
\int_{\mathbb{R}^d} K(\mathbf{x}_{k+1},t_{k+1}|\mathbf{x}_k,t_k)
K(\mathbf{x}_k,t_k|\mathbf{x}_{k-1},t_{k-1}) \exp\left( - \epsilon
\mathbf{x}_k^2 \right) d \mathbf{x}_{t_k} \ee \be \nonumber =
K(\mathbf{x}_{k+1},t_{k+1}|\mathbf{x}_{k-1},t_{k-1}). \ee This
property is the analogue of the
Einstein-Smoluchowski-Chapman-Kolmogorov equation in
the theory of Brownian motion. This property is essential if
$\psi_\alpha$ is to depend only on the homogeneous event $\alpha$ and
not its representation in terms of the pair $(\textbf{t},
\bm{\alpha})$ and we assume it holds for all propagators henceforth.


Having defined restricted evolution according to a homogeneous event
we now define it for all events in $\EA$. Let $\alpha$ be an
event given by \be \alpha = \bigcup_{k=1}^M
\alpha_H^k, \ee with the $\alpha_H^k$ ($k=1,\ldots,M$) a finite
collection of mutually disjoint homogeneous events. We define
$\psi_\alpha$ as the sum \be \label{eq:RestrictedEvolInhomo}
\psi_{\alpha} := \sum_{k=1}^M \psi_{\alpha_H^k}. \ee If the propagator
satisfies the composition property \eref{eq:ESCK} this doesn't depend
on the representation of $\alpha$ as a union of homogeneous events.

For two
events $\alpha, \beta \in \EA$
and an initial normalised vector $\psi \in
\mathcal{L}^2(\mathbb{R}^d)$ one can show that
the decoherence functional (\ref{dcf4qm}) on
$\EA \times \EA$ is equal to the inner product
\be \label{eq:StandardDF} D(\alpha,\beta) :=
\langle \psi_\alpha, \psi_\beta\rangle_0 \,,
\ee
by using the familiar expression for the propagator $K$ as
a path integral
\be\label{eq:pathint}
K(\mathbf{x}_2,t_2|\mathbf{x}_1,t_1) = \int [d\gamma] e^{i S[\gamma]}
\ee
where the integral is over all paths $\gamma$ which begin at $\mathbf{x}_1$
at $t_1$ and end at $\mathbf{x}_2$ at $t_2$.

\subsection{Isomorphism} \label{ParticleIsomorphism}

Henceforth we assume the initial
state $\psi \in \mathcal{L}^2(\mathbb{R}^d)$ has unit norm,
the decoherence functional for events in $\EA$ is given by
the propagator $K$ as described in
section \ref{ParticleDecoherence}, the
spaces $H_1$, $H_2$ are defined as in sections \ref{sec:h1} and \ref{sec:h2}.
We will find conditions on the initial state and
propagator that ensure the History Hilbert space $(H_2,\IP_2)$ is
isomorphic to $(\L2, \IP)$.

It will prove useful to define a map $f_0 : H_1 \to \L2$ given by \be
\label{eq:DefF0} f_0(u):= \sum_{\alpha \in \EA} u(\alpha) \left[
\psi_\alpha \right], \ee for all $u \in H_1$. The sum is well-defined
since $u(\alpha)$ is only non-zero for a finite number of events $\alpha
\in \EA$. This map $f_0$ is linear and for all $u, v \in H_1$, we have
\be
\label{eq:ParticleF0InnerProduct} \langle f_0(u),f_0(v)\rangle =
\langle u, v \rangle_1. \ee

Since the map $f_0$ is linear and preserves the inner products in
$H_1$ and $\L2$ it maps a Cauchy sequence, $\{u_n\}$ of
elements of $H_1$ to a
Cauchy sequence in $\L2$. Since $\L2$ is complete this sequence has a
limit and it is this limit we assign as the image of our candidate
isomorphism, $f : H_2 \to \L2$ defined by: \be \label{eq:DefF} f([u_n]_1):=
\lim_{ n\to \infty} f_0(u_n). \ee

The map $f$ is linear and well-defined, independent of which
representative, $\{u_n\}$ of the $[u_n]_1$ equivalence class is used
in the definition above.

Using \eref{eq:ParticleF0InnerProduct} and the continuity of the
$\IP$ inner product \cite[Lemma 3.2-2]{Kreyszig} we have \be \nonumber
\langle f([u_n]_1), f([v_n]_1) \rangle:= \langle \lim_{n\to \infty}
f_0(u_n),\lim_{m\to \infty} f_0(v_m) \rangle \ee \be
\label{eq:ParticleFInnerProduct} = \lim_{n\to\infty} \langle f_0(u_n),
f_0(v_n) \rangle = \lim_{n\to\infty}\langle u_n , v_n \rangle_1 =:
\langle [u_n]_1, [v_n]_1 \rangle_2 .\ee



By Lemma \ref{lem:OneToOne}, since $f$ is linear and satisfies
\eref{eq:ParticleFInnerProduct}, it is one-to-one.  We can now
state our main theorem:

\begin{theorem}[Onto] \label{the:MainTheorem}
Let the propagator $K(\mathbf{x}_T,T|\mathbf{x}_0,0)$ be
continuous as a function of $(\mathbf{x}_T,\mathbf{x}_0) \in
\mathbb{R}^{2d}$ and such that
for each $\mathbf{x}_T$, $\exists \mathbf{x}_0$
with  $K(\mathbf{x}_T,T|\mathbf{x}_0,0)$ non-zero.
Then the map $f$ defined by \eref{eq:DefF} is onto.
\end{theorem}

To prove Theorem \ref{the:MainTheorem} we follow a strategy suggested
by the proof of Theorem 1: we want to show,
roughly, that every final position
can be reached by a history of nonzero amplitude. The implementation
of the strategy is more complicated than in the finite case and
will proceed by establishing a series of Lemmas.

\begin{lemma} \label{lem:Continuous} Let the
propagator $K(\mathbf{x}_T,T|\mathbf{x}_0,0)$
be continuous as a function of $(\mathbf{x}_T,\mathbf{x}_0)
 \in \mathbb{R}^{2d}$. Let $\psi \in
\mathcal{L}^2(\mathbb{R}^d)$ be the initial state.
 Let $A \subset \mathbb{R}^d$ be a compact measurable set
and $\alpha$ be the homogeneous event represented by
$\mathbf{t} = (0,T), \bm{\alpha}
= (A,\mathbb{R}^d)$. Then
\be \psi_\alpha(\mathbf{x}_T,T):=
\int_A K(\mathbf{x}_T,T|\mathbf{x}_0,0) \psi(\mathbf{x}_0)
d\mathbf{x}_0,
\ee
is continuous as a function of $\mathbf{x}_T \in
\mathbb{R}^d$.

\begin{proof} Fix a position $\mathbf{x}_T \in \mathbb{R}^d$ at
the final time. Let $C$ be the closed unit ball
centred at $\mathbf{x}_T$. By
assumption, $K(\mathbf{x}_T,T|\mathbf{x}_0,0)$ is continuous (as a
function of $(\mathbf{x}_T,\mathbf{x}_0) \in \mathbb{R}^{2d}$) so, by
the Heine-Cantor theorem, it is uniformly continuous (as a function of
$(\mathbf{x}_T,\mathbf{x}_0)$) on the compact set $C \times A \subset
\mathbb{R}^{2d}$. This means for any $\epsilon > 0$ there exists
$\delta > 0$ such that for $(\mathbf{x}_T,\mathbf{x}_0),
(\mathbf{x}_T',\mathbf{x}_0') \in C \times A$ we have \be\fl
\sqrt{|\mathbf{x}_T-\mathbf{x}_T'|^2 + |\mathbf{x}_0-\mathbf{x}_0'|^2}
< \delta \Rightarrow \left| K(\mathbf{x}_T,T|\mathbf{x}_0,0) -
K(\mathbf{x}'_T,T|\mathbf{x}'_0,0) \right| < \epsilon.\ee

In particular if $\mathbf{x}_0 = \mathbf{x}_0'$ and
$|\mathbf{x}_T - \mathbf{x}_T'| < \delta < 1$ then
\begin{equation}
\Big|
K(\mathbf{x}_T,T|\mathbf{x}_0,0) -
K(\mathbf{x}_T',T|\mathbf{x}_0,0)\Big| < \epsilon \quad \quad\forall \mathbf{x}_0 \in A\,.
\end{equation}

So for $|\mathbf{x}_T - \mathbf{x}_T'| < \delta < 1$ we have
\begin{eqnarray*}
|\psi_\alpha&(\mathbf{x}_T,T) - \psi_\alpha(\mathbf{x}_T',T)|\\
             &:= \left|
\int_A \Big( K(\mathbf{x}_T,T|\mathbf{x}_0,0) -
K(\mathbf{x}'_T,T|\mathbf{x}_0,0) \Big) \psi(\mathbf{x}_0)
d\mathbf{x}_0 \right|\\
&\leq \left(\int_A \Big|
K(\mathbf{x}_T,T|\mathbf{x}_0,0) -
K(\mathbf{x}_T',T|\mathbf{x}_0,0)\Big|^2
d\mathbf{x}_0\right)^\frac{1}{2} \left(\int_A |\psi(\mathbf{x}_0)|^2
d\mathbf{x}_0\right)^\frac{1}{2}\\
 & < \epsilon |A|
\end{eqnarray*}
 where we have used the Cauchy-Schwarz
inequality and the normalisation of
$\psi$ and $|A|$ is the Lebesgue measure of $A$.
$|A|$ is finite so, since
$\epsilon$ is arbitrary, $\psi_\alpha(\mathbf{x}_T,T)$ is continuous
at $\mathbf{x}_T$. This holds for any $\mathbf{x}_T \in \mathbb{R}^d$.
\end{proof}
\end{lemma}
\begin{lemma} \label{lem:NonZero} Let the propagator
$K(\mathbf{x}_T,T|\mathbf{x}_0,0)$ be continuous as
a function of $(\mathbf{x}_T,\mathbf{x}_0)$ and be  such that
for each $\mathbf{x}_T$, $\exists \mathbf{x}_0$ s.t.
$K(\mathbf{x}_T,T|\mathbf{x}_0,0)$ is non-zero.
 Then for any
point $\mathbf{x}_T \in \mathbb{R}^d$ at the truncation time $t=T$
there exists a compact measurable set $A \subset \mathbb{R}^d$
(depending on $\mathbf{x}_T$) such that the homogeneous event $\alpha$
represented by $\mathbf{t} = (0,T), \bm{\alpha} = (A,\mathbb{R}^d)$
satisfies \be \label{eq:ParicleNonZero}
\psi_{\alpha}(\mathbf{x}_T,T):=\int_A K(\mathbf{x}_T,T|\mathbf{x}_0,0)
\psi(\mathbf{x}_0) d\mathbf{x}_0 \neq 0.\ee

\begin{proof}
The proof relies on Lebesgue's Differentiation Theorem
\cite[p100]{WheeZyg} which states that if $G : \mathbb{R}^d \to
\mathbb{C}$ is an integrable function then \be \label{eq:LDTheorem}
G(\mathbf{x}) = \lim_{B \to \mathbf{x}} \frac{\int_B G(\mathbf{x}')
d\mathbf{x}'}{|B|},\ee for almost all $\mathbf{x} \in
\mathbb{R}^d$. Here $B$ is an $d$-dimensional ball centred on
$\mathbf{x}$ which contracts to $\mathbf{x}$ in the limit
and $|B|$ is its Lebesgue measure.

Aiming for a contradiction we assume that \be
\int_A
K(\mathbf{x}_T,T|\mathbf{x}',0) \psi(\mathbf{x}') d\mathbf{x}' =
0,\ee for all compact measurable sets $A \subset \mathbb{R}^d$.

Taking $A$ to be a sequence of closed balls contracting to an
arbitrary point $\mathbf{x} \in \mathbb{R}^d$ at the initial time
then \eref{eq:LDTheorem} gives \be K(\mathbf{x}_T,T|\mathbf{x},0)
\psi(\mathbf{x}) = \lim_{A \to \mathbf{x}} \frac{\int_{A}
K(\mathbf{x}_T,T|\mathbf{x}',0) \psi(\mathbf{x}')
d\mathbf{x}'}{|A|} = 0,\ee
for almost all $\mathbf{x} \in \mathbb{R}^d$.
This is a contradiction
since $K$ is continuous and
$\exists \mathbf{x}_0$ with $K(\mathbf{x}_T,T|\mathbf{x}_0,0) \ne0$
so there is
a compact set
containing $\mathbf{x}_0$ on which
$K(\mathbf{x}_T,T|\mathbf{x},0) \ne0$.

\end{proof}
\end{lemma}

\begin{lemma} \label{lem:FiniteProj} Let the propagator
$K(\mathbf{x}_T,T|\mathbf{x}_0,0)$
satisfy the conditions of Lemma \ref{lem:NonZero}.
Then, for any point $\mathbf{x}_T \in \mathbb{R}^d$
at the truncation time there exists a homogeneous event $\alpha$
represented by $\mathbf{t} = (0,T), \bm{\alpha}
= (A,B)$ (with $A \subset \mathbb{R}^d$ a compact measurable set and
$B \subset \mathbb{R}^d$ an open ball centred on $\mathbf{x}_T$) and a
strictly positive real number $P$ such that $\psi_\alpha$ is uniformly
continuous in $B$ and $|\psi_\alpha(\mathbf{x},T)| > P$ for all
$\mathbf{x} \in B$.
\begin{proof}
By Lemmas \ref{lem:Continuous} and \ref{lem:NonZero}, there
exists a compact measurable set $A \subset \mathbb{R}^d$ such that,
for the homogeneous event $\beta$ represented by $\bm{\beta} =
(A,\mathbb{R}^d)$ the function $\psi_\beta(\mathbf{x},T)$ is
continuous for all $\mathbf{x} \in \mathbb{R}^d$ and satisfies
$\psi_\beta(\mathbf{x}_T,T) \neq 0$.

This implies there exists $\delta > 0$ such that \be
\label{eq:DeltaBall} | \mathbf{x} - \mathbf{x}_T | < \delta \Rightarrow
|\psi_\beta(\mathbf{x},T) - \psi_\beta(\mathbf{x}_T,T)| <
\frac{|\psi_\beta(\mathbf{x}_T,T)|}{2}. \ee Let $B$ be the open ball
of radius $\delta$ centred on $\mathbf{x}_T$. Setting $P =
|\psi_\beta(\mathbf{x}_T,T)|/2 > 0$ we see that $\mathbf{x} \in B$
implies $|\psi_\beta(\mathbf{x},T)| > P > 0$.

Since $\psi_\beta(\mathbf{x},T)$ is continuous it is uniformly
continuous in any compact set and therefore any subset of a compact
set. It is thus uniformly continuous in $B$. For $\bm{\alpha} :=
(A,B)$ we then have $\psi_\alpha = \chi_B \psi_\beta$
($\chi_B$ is the characteristic function of $B$)
and the result
follows.
\end{proof}
\end{lemma}

The next lemma is the heart of the proof.

\begin{lemma} \label{lem:ParticleIntervals}
 Let the propagator
$K(\mathbf{x}_T,T|\mathbf{x}_0,0)$
satisfy the conditions of Lemmas \ref{lem:Continuous} and \ref{lem:NonZero}.
 Let $I$ be a compact $d$-interval with positive measure $|I|>0$ at the
truncation time. Then for any $\epsilon > 0$ there exists a vector $u
\in H_1$ such that
\be || [\chi_I] - f_0(u) || < \epsilon\,.
\ee

\begin{proof}
Let $\epsilon >0$.
For any $\mathbf{x} \in I$ there exists, by Lemma
\ref{lem:FiniteProj}, a homogeneous event $\alpha_\mathbf{x}$
represented by $\bm{\alpha_\mathbf{x}} = (A_\mathbf{x}, B_\mathbf{x})$
(with $B_\mathbf{x}$ an open ball centred on $\mathbf{x}$) and a real
number $P_\mathbf{x} > 0$ such that $\psi_{\alpha_{\mathbf{x}}}$ is
uniformly continuous in $B_\mathbf{x}$ and
$|\psi_{\alpha_\mathbf{x}}(\mathbf{x}',T)| > P_\mathbf{x}$ for all
$\mathbf{x}' \in B$.

The collection of $B_\mathbf{x}$, taken for all $\mathbf{x} \in I$,
form an open cover of $I$, which, since $I$ is compact, admits a
finite subcover labelled by $\{\mathbf{x}_i \in I\, |\, i = 1 \dots
N\}$. Define $A_i := A_{\mathbf{x}_i}$,
$B_i := B_{\mathbf{x}_i}$, $\alpha_i:= \alpha_{\mathbf{x}_i}$
and $P_i := P_{\mathbf{x}_i}$.

Each $B_i$ will now be ``cut up'' into finitely many disjoint sets,
$D_{ij}$, over which the $\psi_{\alpha_i}$ functions vary by only
``small amounts''. The first step toward this is to form a finite
number of $N$ mutually disjoint sets $C_i \subseteq B_i$ given by \be
C_1 := B_1 \cap I,\quad C_i := (B_i \cap I) \setminus
\bigcup_{j=1}^{i-1} C_j \quad (i=2,\ldots,N),\ee and such that \be I =
\bigcup_{i=1}^N C_i. \ee
Without loss of generality we assume the $C_i$ are non-empty.

Each function $\psi_{\alpha_i}$ is uniformly continuous in $C_i$ and
satisfies $|\psi_{\alpha_i}(\mathbf{x},T)| > P_i$ for all $\mathbf{x}
\in C_i$ for some strictly positive $P_i \in \mathbb{R}$.

Let $P > 0$ be the minimum value of the $P_i$ and let $\delta_i > 0$
($i=1,\ldots,N$) be chosen such that \be \label{eq:FirstInequal}
|\mathbf{x} - \mathbf{y}| < \delta_i \Rightarrow
|\psi_{\alpha_i}(\mathbf{x},T) - \psi_{\alpha_i}(\mathbf{y},T)| <
\frac{\epsilon P}{\sqrt{|I|}}, \ee for all $\mathbf{x}, \mathbf{y}
\in C_i$.

Letting $\delta > 0$ be the minimum of the $\delta_i$ now subdivide
each $C_i$ into a finite number, $M_i$, of non-empty disjoint sets
$D_{ij}$ ($i=1,\ldots,N; j=1,\ldots,M_i$) such that \be
\bigcup_{j=1}^{M_i} D_{ij} = C_i \quad \textrm{and} \quad
\mathbf{x},\mathbf{y} \in D_{ij} \Rightarrow |\mathbf{x} - \mathbf{y}| <
\delta. \ee

If we arbitrarily choose points $\mathbf{x}_{ij} \in D_{ij}$ the
$D_{ij}$ sets are ``small enough'' that, by \eref{eq:FirstInequal},
$|\psi_{\alpha_i}(\mathbf{x}_{ij},T) - \psi_{\alpha_i}(\mathbf{x},T)|
< \epsilon P/\sqrt{|I|}$ for all $\mathbf{x} \in D_{ij}$. Defining
homogeneous events $\alpha_{ij}$ to be represented by
$\bm{\alpha_{ij}}:=(A_i, D_{ij})$ therefore gives \be
\label{eq:SecondInequal} \fl\left| 1 -
\frac{\psi_{\alpha_{ij}}(\mathbf{x},T)}{\psi_{\alpha_{ij}}
(\mathbf{x}_{ij},T)}\right|
= \frac{|\psi_{\alpha_{ij}}(\mathbf{x}_{ij},T) -
\psi_{\alpha_{ij}}(\mathbf{x},T)|}{|\psi_{\alpha_{ij}}(\mathbf{x}_{ij},T)|}
< \frac{1}{|\psi_{\alpha_{ij}}(\mathbf{x}_{ij},T)|} \frac{\epsilon
P}{\sqrt{|I|}} < \frac{\epsilon}{\sqrt{|I|}},\ee for all
$\mathbf{x}\in D_{ij}$ where we note
$|\psi_{\alpha_{ij}}(\mathbf{x}_{ij})| > P > 0$.

Now define a $H_1$ vector by

\be \fl u(x) := \left\{\begin{array}{rl}
1/\psi_{\alpha_{ij}}(\mathbf{x}_{ij},T) & \textrm{if } x =
\alpha_{ij} \textrm{ for } i=1,\ldots,N; j=1,\ldots,M_i \\ 0 & \textrm{otherwise.}\end{array} \right.\ee

This is a well-defined vector in $H_1$ since there are only a finite
number of events $x=\alpha_{ij}$ on which $u(x)$ is non-zero.  We now
compute

\be || [\chi_I] - f_0(u)||^2 = \int_{\mathbb{R}^d} \left|
\chi_I(\mathbf{x}) - \sum_{i=1}^N \sum_{j=1}^{M_i}
\frac{\psi_{\alpha_{ij}}(\mathbf{x},T)}{\psi_{\alpha_{ij}}(\mathbf{x}_{ij},T)}
\right|^2 d\mathbf{x}\ee \be \nonumber = \sum_{i=1}^N \sum_{j=1}^{M_i}
\int_{D_{ij}} \left| 1 -
\frac{\psi_{\alpha_{ij}}(\mathbf{x},T)}{\psi_{\alpha_{ij}}(\mathbf{x}_{ij},T)}
\right|^2 d\mathbf{x} < \sum_{i=1}^N \sum_{j=1}^{M_i} \int_{D_{ij}}
\frac{\epsilon^2}{|I|} d\mathbf{x} = \epsilon^2,\ee where we have
used \eref{eq:SecondInequal}, the disjointness of the $D_{ij}$ and
\be \sum_{i=1}^N \sum_{j=1}^{M_i} \int_{D_{ij}} d\mathbf{x} =
|I|.\ee We have thus constructed $u \in H_1$ such that \be
||[\chi_{I}] - f_0(u) || < \epsilon. \ee
\end{proof}
\end{lemma}
We can now prove Theorem \ref{the:MainTheorem}.

\begin{proof} (of Theorem \ref{the:MainTheorem})

A \emph{step function} on $\mathbb{R}^d$ is a function $S :
\mathbb{R}^d \to \mathbb{C}$ that is a finite linear
combination of characteristic functions of compact $d$-intervals.

Let $[\phi] \in \L2$ be the element we wish to map to. We assume
$[\phi] \neq 0$ for otherwise the zero vector in $H_2$ would satisfy
$f(0) = [\phi]$. Let $\{[S_n]\}$ be a sequence of $\L2$ vectors,
where the $S_n$ are step functions
that are not identically zero,
such that \be \label{eq:OntoInequal} || [\phi] -
[S_n] || < \frac{1}{2 n}, \ee for each positive integer $n$. Such a
sequence $\{[S_n]\}$ exists since the step functions are dense
in $\mathcal{L}^2(\mathbb{R}^d)$ \cite[p133]{WheeZyg}.

For each step function $S_n$ we can (non-uniquely) decompose it as \be
S_n = \sum_{i=1}^{N_n} s_{n,i} {\chi}_{I_{n,i}},\ee for a finite
collection of $N_n \geq 1$ non-zero complex numbers
$s_{n,i}$ and mutually disjoint compact $d$-intervals $I_{n,i}$.
Define $M_n > 0$ to be the maximum value of $|s_{n,i}|$
($i=1,\ldots,N_n$).

By Lemma \ref{lem:ParticleIntervals}, for each $n = 1,2,\ldots$ and
each $i=1,\ldots,N_n$ there exists a vector $u_{n,i} \in H_1$ such
that \be || [\chi_{I_{n,i}}] - f_0(u_{n,i}) || < \frac{1}{2 n N_n
M_n}. \ee Defining $u_n \in H_1$ by \be u_n := \sum_{i=1}^{N_n}
s_{n,i} u_{n,i}, \ee we see that \be \fl ||[S_n] - f_0(u_n)|| \leq
\sum_{i=1}^{N_n} |s_{n,i}| || [\chi_{I_{n,i}}] - f_0(u_{n,i}) || <
\sum_{i=1}^{N_n} \frac{|s_{n,i}|}{2 n N_n M_n} < \sum_{i=1}^{N_n}
\frac{1}{2 n N_n} = \frac{1}{2 n}. \ee This, together with
\eref{eq:OntoInequal}, implies \be || [\phi] - f_0(u_n) || <
\frac{1}{n}, \ee i.e. $f_0(u_n)$ is a Cauchy sequence converging to
$[\phi]$. Since $f_0$ preserves the inner product
this means $\{ u_n \}$ is a Cauchy sequence of $H_1$ elements such
that $f([u_n]_1) = [\phi]$.   $[\phi] \in \L2$ was arbitrary
so the map $f$ is onto.
\end{proof}

Theorem \ref{the:MainTheorem} gives sufficient conditions
on the propagator for the History Hilbert space to be
isomorphic to $\L2$ \emph{for any initial state}. If the
initial state itself satisfies certain conditions, then the
conditions on the propagator can be relaxed. For example,
if the initial state, $\psi$, is everywhere nonzero, then
even a trivial evolution with a delta-function propagator
will suffice to make the History Hilbert space
isomorphic to $\L2$.

\subsection{Examples}
\label{sec:ParticleExamples}
We now look at
examples for which the propagator is known explicitly. The expressions
for the propagators are taken from \cite{Handbook}.

For a free particle of mass $m$ in $d$ dimensions the Lagrangian is
\be L = \frac{m}{2} \dot{\mathbf{x}}^2.\ee The propagator is given by
\be K(\mathbf{x}',t'|\mathbf{x},t) = \left( \frac{m}{2 \pi i \hbar
(t'-t)} \right)^{d/2} \exp\left[ \frac{im}{2 \hbar (t'-t)}
(\mathbf{x}'-\mathbf{x})^2 \right]. \ee For a charged particle (with
mass $m$ and charge $e$) in a constant vector potential $\mathbf{A}$
the Lagrangian is \be L = \frac{m}{2} \dot{\mathbf{x}}^2 + e\mathbf{A}
\cdot \dot{\mathbf{x}}. \ee The propagator is given by \be\fl
K(\mathbf{x}',t'|\mathbf{x},t) = \left( \frac{m}{2 \pi i \hbar (t'-t)}
\right)^{d/2} \exp\left[ \frac{im}{2 \hbar (t'-t)}
(\mathbf{x}'-\mathbf{x})^2 + \frac{i e \mathbf{A}}{\hbar} \cdot
(\mathbf{x}' - \mathbf{x})\right]. \ee Both of these propagators
satisfy the conditions for Theorem \ref{the:MainTheorem}.
Since the system with constant vector potential is gauge
equivalent to the free particle, the theorem is bound to
hold for both or neither.

A particle of mass $m$ in a simple harmonic
oscillator potential of period $2 \pi/\omega$ in one spatial
dimension has Lagrangian
\be L = \frac{m}{2} \dot{x}^2 -
\frac{m\omega^2}{2} x^2\,. \ee
Defining, $\Delta t := t' - t$, the
propagator is
\be \fl K(x',t'|x,t) = \left( \frac{m\omega}{2 \pi i \hbar
\sin(\omega \Delta t)} \right)^{1/2} \exp\left[ -\frac{m\omega}{2 i
\hbar}\left[ (x'^2+x^2) \cot(\omega \Delta t) - 2 \frac{x
x'}{\sin(\omega \Delta t)} \right] \right], \ee
if $\Delta t \neq M \pi / \omega$ for integer $M$.

If $\Delta t = M \pi/\omega$ for integer $M$ we have
\be K(x',t'|x,t) = e^{(-i M \pi/2)}\delta(x' - (-1)^M x). \ee

Clearly the propagator
fulfils the conditions for Theorem \ref{the:MainTheorem} if the
truncation time $T$ is not equal to $M \pi/\omega$ for integer $M$.
More care is needed if the truncation time is an integer multiple of
$\pi/\omega$.

If $T = M \pi/\omega$ for integer $M$ then the
propagator does not fulfil the conditions for Theorem
\ref{the:MainTheorem}. In this case we cannot use only two-time events
to demonstrate the isomorphism. The Hilbert spaces are still
isomorphic, however, as can be seen by using three-time homogeneous
events $\alpha$ represented by $\bm{\alpha}=(\mathbb{R},\alpha_{t_2},\alpha_T)$ in which the set
at time $t_1=0$ is $\mathbb{R}$ and such that $T-t_2$ is not an
integer multiple of $\pi/\omega$. Evolving the initial state according
to these events is equivalent to unrestrictedly evolving the initial
state from $t_1=0$ to $t_2 > 0$. The state at time $t_2$ can then be
viewed as the ``initial state'' for two-time homogeneous events represented by
$(\alpha_{t_2},\alpha_T)$. The conditions for Theorem
\ref{the:MainTheorem} are met by $K(x_T,T|x_{t_2},t_2)$ so the theorem
can be applied and the isomorphism demonstrated. These ideas can
similarly be applied to the simple harmonic oscillator in $d$
dimensions.

\subsection{Particle with an infinite potential barrier} \label{ParticlePotential}

Consider a physical system of a non-relativistic particle in one
dimension restricted to the positive halfline $\mathbb{R}^+ = \{x \in
\mathbb{R} | x > 0 \}$ by an infinite potential barrier.

The Hilbert space for this system is $L^2(\mathbb{R}^+)$ which we
define as a vector subspace of $L^2(\mathbb{R})$:

\be L^2(\mathbb{R}^+):= \{ [\psi] \in L^2(\mathbb{R}) : \psi(x) = 0
\textrm{ for } x \leq 0\}. \ee

The sample space, $\Omega$, and event algebra, $\EA$, for this system
will be the same as for a particle in 1 dimension. The difference is
that, when defining the decoherence functional we now use an initial
vector $\psi \in \mathcal{L}^2(\mathbb{R})$ such that $\psi(x) = 0$
for $x \leq 0$ and a propagator defined by \cite[p40]{Schulman}: \be
K(x',t'|x,t) = \chi_{\mathbb{R}^+}(x') \chi_{\mathbb{R}^+}(x)
\left(\frac{m}{2\pi i \hbar (t'-t)}\right)^{1/2} \ee \be \nonumber
\times \left[ \exp\left[\frac{im(x'-x)^2}{2\hbar(t'-t)}\right] -
\exp\left[\frac{i m(x' + x)^2}{2\hbar(t'-t)}\right] \right], \ee where
$m$ is the mass of the particle.

This propagator does not satisfy the conditions of Theorem
\ref{the:MainTheorem}---it is continuous as a function of $(x,x') \in
\mathbb{R}^2$ but is zero for $x \leq 0$ or $x' \leq 0$. It is not
surprising therefore that the map $f : H_2 \to L^2(\mathbb{R})$ defined by \eref{eq:DefF} is not an isomorphism
with this event algebra and decoherence functional, namely because $f$ only gives vectors $[\psi] \in
L^2(\mathbb{R}^+)$ as expected.

It is possible to show, by using the same methods used in the
isomorphism proof for a particle in $d$ dimensions, that the
History Hilbert space for this event algebra and decoherence functional
is isomorphic to $L^2(\mathbb{R}^+)$.

\subsection{Infinite times} \label{InfTime}

In the preceding sections we assumed a finite time interval
both in the finite configuration space case and the quantum mechanics
case. We can extend the analysis to cover all times to the
future of the initial time,
$t\in [0, \infty)$.
We will describe how to do this
in the quantum mechanics case; the extension can be applied,
\emph{mutatis mutandis}, to the
finite configuration space case.

The sample space $\Omega$ is now the set of continuous real functions
on $[0,\infty)$. The homogeneous events, $\alpha$, are defined
as before as represented by a positive integer $N\ge 1$, an $N$-tuple of
times $\mathbf{t}= (t_1=0, t_2, \dots t_N)$ and an $N$-tuple of
measurable subsets of
$\mathbb{R}^d$, $\mathbf{\alpha} = (\alpha_1, \alpha_2, \dots, \alpha_N)$
such that either $\alpha_k$ or $\alpha_k^c$ is bounded.
Now, however,
there is no truncation time and therefore no restriction on the times $t_k$, they can be
arbitrarily large.
The event algebra, $\EA$, is the set of finite unions of the homogeneous
events. Since there is no common truncation time $T$, the restricted evolution
of the initial state with respect to a
homogeneous event, as defined by \ref{eq:RestrictedEvolHomo},
results in a state defined at a time, $t_N$, that depends on the
event. Such states cannot be added together to define the
restricted evolution
of the initial state with respect to a event which is a
union of disjoint homogeneous events which have different
last times. Instead, we evolve the restricted state back to the
initial time $t=0$, {\emph{i.e.}} we define
\be
\psi_\alpha(\mathbf{x}_0, 0) := \int_{\mathbb{R}^d} \!\!\!\! d\mathbf{x}_N K(\mathbf{x}_0, 0|
\mathbf{x}_N, t_N) \psi_\alpha(\mathbf{x}_N, t_N)
\ee
for each homogeneous event $\alpha$.

We can now work at the initial time. The restricted state
at $t=0$ of
an inhomogeneous event is the sum of the restricted states at $t=0$
of its constituent
disjoint homogeneous events (as in \eref{eq:RestrictedEvolInhomo}).
The decoherence functional is
defined as the inner product of the restricted states
at $t=0$.

The event algebra, $\EA_\infty$, in the infinite time case
contains a subalgebra, $\EA_\infty|_T$, which is canonically
isomorphic
to the event algebra, $\EA_T$ with a truncation time
because each history with a truncation time corresponds to an
event in the infinite time case: the event is the set
of infinite time histories which match the truncated
history. The decoherence functionals on $\EA_\infty|_T$ and $\EA_T$ agree
because the unitary evolution back to the initial time
preserves the inner product. Theorem \ref{the:MainTheorem}
therefore also applies to the semi-infinite time case: if the History Hilbert
space is isomorphic to  $\L2$ with a truncation time, it is isomorphic
without. In the former case it is convenient to construct the
History Hilbert space at the truncation time as we did
and in the latter it
is convenient to consider the History Hilbert space associated with the
initial time, but for the unitary systems we are considering this
is not a real distinction, being akin to working in
the Schr\"odinger or Heisenberg Picture.

\subsection{Mixed states}

The conjecture made at the end of section 2 was that where both standard
and History Hilbert spaces exist, generically they are isomorphic if the
decoherence functional encodes a pure initial state.
If in contrast the initial state is a statistical mixture then the
decoherence functional is a convex combination of
decoherence functionals, and then the History Hilbert space can be
bigger than the standard Hilbert space.
Indeed, we make note of the following expectations for the case of a
finite configuration space.
If the initial state is a density matrix of rank $r_i$
%
%
%
then the History Hilbert space is generically
the direct sum of $r_i$ copies of the standard
Hilbert space $\mathbb{C}^n$. 
Even more generally,
if there is also a final density matrix of rank $r_f$
then the History Hilbert space is
the direct sum of $r_i$ copies of $\mathbb{C}^{r_f}$.
\cite{MarieDavid}

\section{Discussion} \label{Conclusions}

\def\HS{H}

If histories-based formulations of quantum mechanics are nearer to the
truth than state- and operator-based formulations, and in particular if
something like Quantum Measure Theory is the right framework for a
theory of quantum gravity, then there is no particular reason why one
should expect Hilbert spaces to be part of physics at a fundamental
level.

Indeed, in histories formulations which assume only plain,
``weak'' positivity
(and in which, therefore, no Hilbert space arises),
certain kinds of devices can in principle exist
that are not possible within ordinary quantum mechanics,
and this could be regarded as desirable.
For example non-signaling correlations of the ``PR box'' type become
possible \cite{jointDCF}.

Nor does reference to a Hilbert space seem to be needed for
interpretive reasons.  On the contrary, attempts to overcome the
``operationalist'' bias of the so called Copenhagen interpretation
tend to lead in the opposite direction, away from state-vectors and
toward histories and the associated events \cite{QuantumMeasure2}.

Thus, it seems hard to argue on principle that a Hilbert space is needed.
On the other hand,
there do exist good reasons to regard strong positivity
as more natural than weak positivity.
First, it is mathematically much simpler than weak positivity, whence
more amenable to being verified and worked with \cite{SorkinWalk}.
(Not that its definition is any simpler, but that it comprises, apparently, far
fewer independent conditions.)
Second, strong positivity is preserved under composition of subsystems,
whereas the obvious ``product measure'' of two weakly positive
quantal measures is not in general positive at all.
And third --- at a technical level ---
the histories hilbert space to which
strong positivity leads has already proven to be
useful in certain applications
\cite{jointDCF,q-covers},
while there are also indications that the map taking events
$\alpha\in\EA$ to vectors in $\HS$
could be of aid in the effort to extend the decoherence functional
from $\EA$ to a larger fragment of the $\sigma$-algebra it generates.

It thus seems appropriate to add strong positivity to the axioms
defining a decoherence functional
(as we have done in this paper),
and from a strongly positive decoherence functional a
histories hilbert space $\HS$
automatically arises.
Once we have it, we can ask whether the histories hilbert space helps us
to make contact with the quantal formalism of standard textbooks.  This
is something that any proposed formulation has to be able to do, and it
is the principal question animating the present paper.  The
positive answer we have obtained is that for the systems we have
studied, the histories hilbert spaces that pertain to them can be
directly identified with the corresponding state-spaces of the
ordinary quantum description.  (The two are ``naturally isomorphic''.)
Thereby an important part of the mathematical apparatus of ordinary
quantum mechanics is recovered quite simply.
This result can be seen as an advance for both Generalised Quantum
Mechanics and Quantum Measure Theory
because the basic underlying structures
 --- histories and decoherence functionals ---
are common to both approaches.
(Strong positivity has not normally been assumed in Generalised Quantum
Mechanics, but there is no reason why it could not be.)

Beyond state-vectors, the other main ingredients of the standard quantum
machinery are the operators representing position, momentum, field
values, ``observables'', and the like.  How might they be derived from
histories?
In the specialized context of unitary, Hamiltonian evolution and the
Schr{\"o}dinger equation,
time-ordered operators can be obtained from functions (``functionals'')
on the sample space $\Omega$ (see \cite{Feynman:1948}),
but whether such a relationship exists in the same generality as the
histories hilbert space itself (that is for any quantum measure theory)
remains to be seen.
An interesting generalization where one does seem able to recover field
 operators from the decoherence functional is that of quantum field
 theory on a causal set \cite{steven}.

The context of this paper has been that of non-relativistic quantum
mechanics, yet people have not yet completely laid
the rigorous mathematical foundations of a
histories framework for this theory.
Nevertheless
the decoherence functional limited to $\EA\times\EA$
is known
(we haven't yet defined calculus but we can calculate
the volume of a pyramid, see footnote 10, page 371 of
\cite{Feynman:1948}),
and this sufficed to demonstrate our main result, that the
History Hilbert space $\HS$ is the
standard Hilbert space.
A key question for the future
that will also be of interpretational significance
is
what the sample
space of histories is.
Is it the set of all continuous trajectories and if so,
exactly how continuous are they?
This is closely related to the
question, can the decoherence functional --- and hence the
quantal measure --- be extended to a larger collection of sets
than $\EA$?  Is that larger collection the whole $\sigma$-algebra
generated by $\EA$ or something smaller?
These questions have been explored by Geroch \cite{GerochPathIntegrals}.
To the extent that they find satisfactory answers,
we will be able to say
that Quantum Mechanics as Quantum Measure Theory is as well-defined
mathematically as the Wiener process.

Be that as it may,
neither Brownian motion nor the quantum mechanics of
nonrelativistic point-particles can lay claim to fundamental status in
present-day physics.
Relativistic quantum field theory comes closer,
but in that context, neither formulation ---
neither path-integrals/histories nor state-vectors-cum-operators ---
enjoys a mathematically rigorous existence.
Instead we have the divergences and other pathologies whose resolution
is commonly anticipated from the side of quantum gravity.
If this expectation is borne out,
the decoherence functional of quantum gravity might actually be
easier to place on a sound mathematical footing than that of the
Hydrogen atom,
because in place of a path-integral over an infinite dimensional
function-space,
we will have something more finitary in nature,
like a summation over a discrete space of histories.\footnote%
{One can already observe such a trend in the theory formulated in
 \cite{steven} of a free scalar field on a causal set $C$ corresponding
 to a bounded spacetime region.  The decoherence functional of the
 theory can be computed and is again given by a double integral
 of the type of (\ref{dcf4qm}).
 Now however, the domain of integration is just $\mathbb{R}^n$
 rather than some infinite-dimensional path-space.
 Moreover, the integrand contains,
 besides
 the expected oscillating phases, damping terms
 that
 lessen by half
 the need for integrating-factors like those in
 \eref{eq:IntConvergenceFactor}.}

In that case the trek back to nonrelativistic quantum mechanics will be
longer, but we expect that the
histories hilbert space defined above will still be an important milestone
along the way.

\ack
The authors would like to thank Chris Isham for helpful criticisms on
the first draft of this paper. We also thank Raquel S. Garcia for
discussions during the early stages of this
work. SJ is supported by
a STFC studentship. FD acknowledges support from  ENRAGE, a Marie Curie Research Training Network contract MRTN-CT-2004-005616
and the Royal Society IJP 2006/R2.
 FD and SJ thank Perimeter Institute for Theoretical Physics for
 hospitality during the writing of this paper.
Research at Perimeter Institute for Theoretical Physics is
supported in part by the Government of Canada through NSERC
and by the Province of Ontario through MRI.

\end{document}